\documentclass[11pt]{iopart}

\usepackage{graphicx}
\usepackage{iopams}  
\usepackage[applemac]{inputenc}

\begin{document}

\title{Gauge fields for ultracold atoms in optical superlattices}
\author{ Fabrice Gerbier and Jean Dalibard}  
\ead{fabrice.gerbier@lkb.ens.fr; jean.dalibard@lkb.ens.fr}
\address{Laboratoire Kastler Brossel, ENS, UPMC-Paris 6, CNRS ; 24 rue Lhomond, 75005 Paris, France}
\date{\today}

%
\begin{abstract}

We present a scheme that produces a strong U(1)-like gauge field on cold atoms confined in a two-dimensional square optical lattice. Our proposal relies on two essential features, a long-lived metastable excited state that exists for alkaline-earth or Ytterbium atoms, and an optical superlattice. As in the proposal by Jaksch and Zoller [New Journal of Physics {\bf 5}, 56 (2003)], laser-assisted tunneling between adjacent sites creates an effective magnetic field. In the tight-binding approximation, the atomic motion is described by the Harper Hamiltonian, with a flux across each lattice plaquette that can realistically take any value between 0 and $\pi$. We show how to take advantage of the superlattice to ensure that each plaquette acquires the same phase, thus simulating a uniform magnetic field. We discuss the observable consequences of the artificial gauge field on non-interacting bosonic and fermionic gases. We also outline how the scheme can be generalized to non-Abelian gauge fields. 
\end{abstract}
%

\section{Introduction}

The fractional quantum Hall (FQH) phases realized by two-dimensional electron gases in very large magnetic ﬁelds are among the most intriguing states of matter (see, for instance,  \cite{girvin1999a}). In such systems, electrons ``bind'' to magnetic vortices, forming strongly correlated phases with striking properties, such as exotic excitations (``anyons'') which obey fractional statistics \cite{stern2007a}. Analogous quantum Hall phases should also arise in cold atomic gases when they are set into fast rotation (see \cite{cooper2008a} and references therein). Due to the mathematical similarity between Coriolis and Lorentz forces, rotating neutral gases are indeed the exact analogue of an assembly of charged particles plunged in a magnetic field. Observing these highly correlated phases is one of the major goals in the field of trapped quantum gases \cite{cooper2008a,bloch2008a}. This goal has, however, not yet been reached, because of the difficulty of communicating the required amount of angular momentum (on the order of $N \hbar$ per atom, with $N$ the number of particles) to the system \cite{schweikhard2004a,bretin2004a}. In practice, the residual static trap anisotropy limits the total angular momentum to much smaller values, for which the rotating gas is well described by a mean field approach \cite{ho2001a}.

Recently alternative schemes to ``simulate'' artificial gauge fields for neutral atoms have been explored using two-dimensional (2D) optical lattices \cite{jaksch2003a,ruostekoski2002a,mueller2004a,sorensen2005a,dudarev2005a,lim2008a,klein2009a}. As they do not involve any mechanical rotation of the system, they should be less sensitive to the imperfection of the trapping potential and thus easier to implement. The guideline for these proposals is the celebrated Harper model \cite{harper1955a,hofstadter1976a}, defined by the two-dimensional (2D) single-particle Hamiltonian,
\begin{eqnarray} 
H_{\rm Harper} &=&-J~\sum_{n,m,\pm }  e^{\pm i 2\pi\alpha m} \hat{c}_{n \pm 1,m}^\dagger \hat{c}_{n,m} +  \hat{c}_{n,m \pm 1}^\dagger \hat{c}_{n,m} + {\rm h.c.} 
\label{eq:hofstadter}
\end{eqnarray}
This tight-binding model was initially introduced \cite{harper1955a} to describe electrons hopping on a square lattice perpendicular to a constant magnetic field $B$. The operator $ \hat{c}_{n,m}^\dagger$ creates a particle at position $(x=n d,y=m d)$, $d$ is the lattice spacing, and $J$ is the tunnel energy to nearest neighbors in the absence of magnetic field. The effect of a vector potential (here in the Landau gauge, ${\bf A}=-B y {\bf e}_{x}$) is included in the so-called Peierls phase factor, $2\pi \alpha m$, where $\alpha=e B d^2/h$ is the flux per unit cell expressed in units of the flux quantum $\phi_{0}=h/e$. The Harper Hamiltonian in Eq.~(\ref{eq:hofstadter}) has been studied extensively in the literature (see, {\it e.g.} \cite{harper1955a,hofstadter1976a,Thouless1982a,kohmoto1989a}). For zero flux, the Harper model reduces to a standard tight-binding model on a square lattice, with a single Bloch band of width $8 J$. For rational fluxes $\alpha=p/q$ (where $p$ and $q$ are integers), the model is still periodic but with larger unit cells of size $qd\times d$. The Bloch band splits into $q$ magnetic sub-bands, and the ground state becomes $q-$fold degenerate. This results in the peculiar self-similar structure  in an energy-magnetic flux diagram known as Hofstadter's butterfly \cite{hofstadter1976a}. This structure exists in a strict sense for infinitely extended systems. For a finite system of size $L$, the details of this structure are washed out on fine scales such that $q > L/d$. 

In their seminal proposition for an implementation of (\ref{eq:hofstadter})  \cite{jaksch2003a}, Jaksch and Zoller (JZ) suggested that the phase $2\pi\alpha$ can be imprinted by a laser beam inducing hopping between adjacent sites (see also \cite{ruostekoski2002a,mueller2004a,sorensen2005a,dudarev2005a,lim2008a,klein2009a} for alternative proposals). A major advantage of this approach is that large flux $\alpha \sim 1$ can be reached under experimentally realistic conditions. For a filling factor with $\sim 1$ atom per site, the quantum Hall regime then becomes reachable in presence of moderate atomic interactions \cite{sorensen2005a,palmer2006a,palmer2008a,hafezi2007a,umucalilar2008a}. Theoretical studies have found that some of these phases were essentially similar to that encountered in the bulk \cite{sorensen2005a,palmer2006a,palmer2008a, umucalilar2008a}, whereas others are without counterpart in the bulk phase \cite{hafezi2008a,moeller2009a}. 

Although the existing proposals for implementing the Harper Hamiltonian with atomic quantum gases open many avenues, they are still challenging from an experimental point of view (for reasons discussed in detail below). In this paper, we extend the JZ proposal and discuss a new scheme based on an optical superlattice \cite{sebbystrabley2006a,foelling2007a} to generate a gauge potential leading to (\ref{eq:hofstadter}). At variance with earlier works where alkali atoms were considered, we propose here to use atoms with a long-lived metastable excited state, such as alkaline-earth \cite{kraft2009a,stellmer2009a,escobar2009a} or Ytterbium \cite{takasu2003a,fukuhara2007b} atoms. For concreteness we will discuss the case of Ytterbium, for which degenerate gases have already been produced for both bosonic and fermionic isotopes \cite{takasu2003a,fukuhara2007b}. We take advantage of this  level structure to alleviate many practical difficulties encountered with alkalis. Importantly, our scheme uses only building blocks which have already been individually demonstrated. 

This paper is organized as follows. We first introduce the system, consisting of atoms with a ground state $g$ and a long-lived metastable excited state $e$ trapped in a two-dimensional (2D) lattice. The optical lattice is tuned near an ``anti-magic'' wavelength where the polarisabilities of $e$ and $g$ are opposite \cite{yi2008a}. The two internal states thus live in spatially separated sublattices. We show that the case $\alpha=1/2$, leading to Dirac points analogous to those appearing in graphene \cite{neto2009a}, can be  achieved rather simply with a single laser beam coupling $e$ and $g$. We then extend our scheme to generate arbitrary values of $\alpha$ using an optical superlattice, at the expense of several lasers operating at different frequencies. Subsequently, we address  the influence of the additional potential that must be added to confine the particles. We investigate its effect by looking at (i) the density of states that determines the density profile of a trapped ideal Fermi gas, (ii) the ground state of the system, relevant for non-interacting bosons. In both cases, we obtain a clear signature for a non-zero value of $\alpha$. We conclude by indicating how the method can be generalized to generate non-Abelian gauge fields \cite{osterloh2005a}.

\begin{figure*}[t]
\centering{\includegraphics{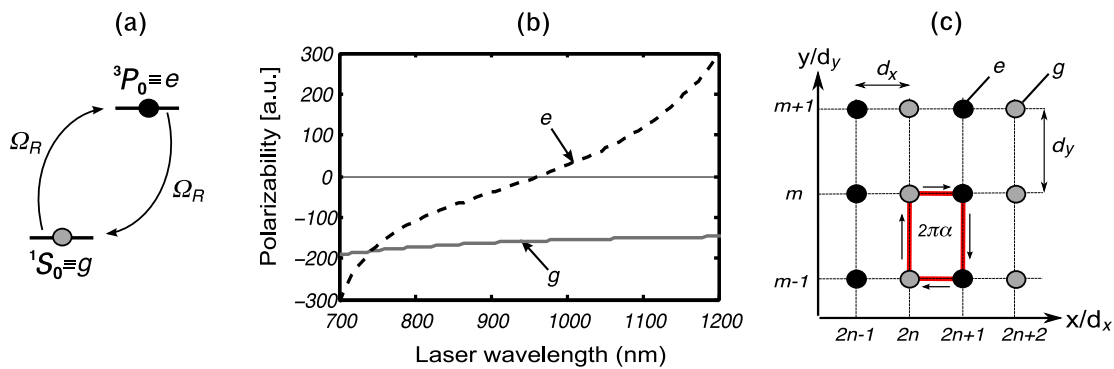}}
\caption{{\bf (a)} Sketch of the level scheme showing the two internal states $g$ and $e$ coherently coupled by a laser beam; {\bf (b)}: Dynamic polarizability $\alpha(\lambda)$ of atoms in the ground (solid line) or excited states (dashed line). The light-shift experienced by the atoms is $\alpha(\lambda)\vert {\bf E}\vert^2/2$, with $\vert {\bf E}\vert$ the electric field strength at wavelength $\lambda$. Two particular wavelengths are of special interest, a ``magic'' wavelength $\lambda_{\rm m}\approx 760~$nm and an ``anti-magic'' wavelength $\lambda_{\rm am}\approx 1.12~\mu$m. {\bf (c)} Sketch of the atomic configuration in a state-dependent optical lattice, showing the two interlaced sublattices for $g$ (grey dots) and $e$ (black dots) atoms. The $y$ lattice is formed by a standing wave at the ``magic'' wavelength and confines both states identically. The $x$ lattice is tuned near the ``anti-magic'' wavelength and confines the two states in two distinct sublattices shifted by $\lambda_{\rm am}/4$.}
\label{fig:YbLevels}
\end{figure*}

\section{Two-electron atoms in state-dependent optical lattices}\label{sec2}

Two-electron atoms, such as Ytterbium and alkaline-earth atoms, generally have a spin-singlet ($^{1}S_{0}\equiv g$) ground state and an extremely long-lived spin triplet ($^{3}P_{0}\equiv e)$ excited state (see Fig.~\ref{fig:YbLevels}a). The very long lifetime of $e$ (around $20~$s for Yb \cite{porsev2004a}) allows one to operate optical atomic clocks at the resonance wavelength $\lambda_0\approx578~$nm between $g$ and $e$ \cite{hong2005a,hoyt2005a,barber2006a}. It has also inspired novel proposals for quantum information processing \cite{reichenbach2007a,stock2008a,gorshkov2009a} or quantum simulation \cite{sato2009a,gorshkov2009b}. States $g$ and $e$ usually react differently to far-off resonance laser light, as they couple optically to many different states. The polarizabilities of the two states calculated using the data from \cite{barberthesis} are shown in Fig.~\ref{fig:YbLevels}b. For laser light tuned at the so-called \emph{magic wavelength} $\lambda_{\rm m}$ ($\sim760~$nm for Yb), both states have the same polarizability and feel the same optical trapping potential \cite{ye2008a}. Central to our proposal is the existence of an ``anti-magic'' wavelength $\lambda_{\rm am}\approx1.12~\mu$m \cite{yi2008a}, for which the polarizabilities of $g$ and $e$ are opposite\footnote{There exists another anti-magic wavelength near $620~$nm, which is unfortunately rather close to a transition connecting to the $^{3}P_{0}$ state ($649~$nm). A lattice at this wavelength is thus likely to suffer from excessive spontaneous emission.}\footnote{A many-electrons calculation \cite{dzuba2009a} using more accurate parameters for the positions and widths of the optical transitions predict the anti-magic wavelengths near $1120.3~$nm and $618.7~$nm [A. Derevianko, private communication (2009)].}. As we will see, this allows to create easily internal-state dependent potentials.

The first building block of our proposal consists in a laser beam at the magic wavelength, forming an horizontal light sheet that freezes the vertical motion of the atoms irrespective of their internal state $e$ or $g$. From now on, we shall focus on the atom dynamics in the horizontal $xy$ plane.The second ingredient consists in an optical standing wave along the $y$ axis, also at the magic wavelength, which generates the same potential $V(y)=-V_0 \cos^2(\pi y/d_{y})$ for both states $g$ and $e$ ($V_{0}>0$). For the third building block, we apply another optical standing wave at the ``anti-magic'' wavelength $\lambda_{\rm am}$ along the $x$ axis, generating the potential $V'(x)=\pm V_1 \cos^2(\pi x/2 d_{x})$, where the $+$ and $-$ signs refer to $e$ and $g$, respectively. The resulting potential $V'(x)+V(y)$ creates two shifted sublattices (see Fig.~\ref{fig:YbLevels}c). Atoms in state $g$ sit at positions ${\bf r}_{g}=(2n,m)$ and atoms in state $e$ at positions ${\bf r}_{e}=(2n+1,m)$, where $n,m$ are integers. Here we take as lengths scales $d_{x}=\lambda_{\rm am}/4$ along $x$, and $d_{y}=\lambda_{\rm m}/2$ along $y$. We write the Wannier functions for the lowest Bloch band as $w_{g}({\bf r}-{\bf r}_{g})$ and $w_{e}({\bf r}-{\bf r}_{e})$. The potential $V_0$ is chosen such that the tunnelling energy $J^{(y)}$ along the vertical lines is significant. On the opposite, we choose $V_1$ large enough to neglect direct tunnelling along the $x$ axis.

Tunnelling along the $x$ axis is induced by a coherent optical coupling between $g$ and $e$. Since atoms in $e$ and $g$ are trapped in separate optical lattices, this coupling induces hopping from one sublattice to the other. Following Jaksch and Zoller \cite{jaksch2003a}, we calculate the effective $g-e$ hopping matrix element between two neighboring sites located at ${\bf r}_{g}=(2n,m)$ and ${\bf r}_{e}={\bf r}_{g}+{\bf b}=(2n+1,m)$ as
\begin{equation}
J_{eg}^{(x)}e^{i {\bf q}\cdot  {\bf r}_{g}}=\frac{\hbar \Omega}{2} e^{i {\bf q}\cdot  {\bf r}_{g}}\int w_{e}^\ast({\bf r} -{\bf b}) e^{i {\bf q}\cdot {\bf r}} w_{g}({\bf r})\;{\rm d}^{2}{\bf r},
\end{equation}
with $\Omega$ and ${\bf q}$ the laser Rabi frequency and wavevector. We suppose that the coupling laser propagates in the $yz$ plane, so that ${\bf q}\cdot  {\bf r}_{g}=2\pi\alpha m$, where $|\alpha|$ can take any value between $0$ and $\lambda_{m}/2\lambda_0$ ($\simeq 0.66$ for Yb) by adjusting the angle between the coupling laser beam and the $z$ axis. 

Before proceeding with further analysis, let us give some numerical estimates for the different physical quantities involved in this experimental configuration. The anti-magic wavelength $\lambda_{\rm am}$ corresponds for Yb to a recoil energy $E_{\rm R}/h=h/(2m\lambda_{{\rm am}}^2)\approx  900~$Hz. Realistic values up to several tens of kHz can be achieved for the depth of the optical lattice potential $V_{1}/h$. In order to avoid excitations to higher-lying Bloch bands, the Rabi frequency $\Omega$ must be small compared to $\Delta/\hbar$, where $\Delta$ is the energy gap between the ground and first excited band. To give numerical estimates, we take $V_{1}=20~E_{\rm R}$, corresponding to a gap $\Delta\approx 8~E_{\rm R}$.  Assuming a moderate Rabi frequency $\Omega= E_{\rm R}/\hbar$, we find a laser-induced tunneling energy $J_{eg}^{(x)}\approx 0.05~E_{\rm R}$, much larger than the residual tunneling energy in each sublattice without change in internal state, $J_{gg}^{(x)}= J_{ee}^{(x)}\approx 2.5\times 10^{-3}E_{\rm R}$. In principle, besides resonant transitions between the fundamental bands of the two sublattices, the coupling laser also induces second-order tunneling processes, where an atom tunnels to a neighboring site in a given sublattice, via a virtual transition to an intermediate excited band in the other sublattice. We have verified that the corresponding tunneling rates were negligible compared to $J_{gg}^{(x)}$ for the range of parameters investigated here. Finally, one can calculate the spontaneous emission and associated heating rates for atoms in each sublattice \cite{gordon1980a}. For lattice depths $V_{0}=V_{1} \sim 30~E_{\rm R}$, we find total heating rates $\gamma_{\rm h}^{(g)} \sim 0.4~$nK/s and $\gamma_{\rm h}^{(e)} \sim 0.9~$nK/s, respectively, or $\gamma_{\rm h}^{(g)}/E_{\rm R} \sim 0.01~$s$^{-1}$ and $\gamma_{\rm h}^{(e)}/E_{\rm R} \sim 0.02~$s$^{-1}$. The heating rates (dominated by the vertical arm at $\lambda_{\rm m}$) are small enough to maintain the atoms in the ground band for several seconds.

\section{State-dependent lattice: Harper Hamiltonian for $\alpha=\frac{1}{2}$}

\begin{figure*}[t!!!]
\centerline{\includegraphics[scale=1]{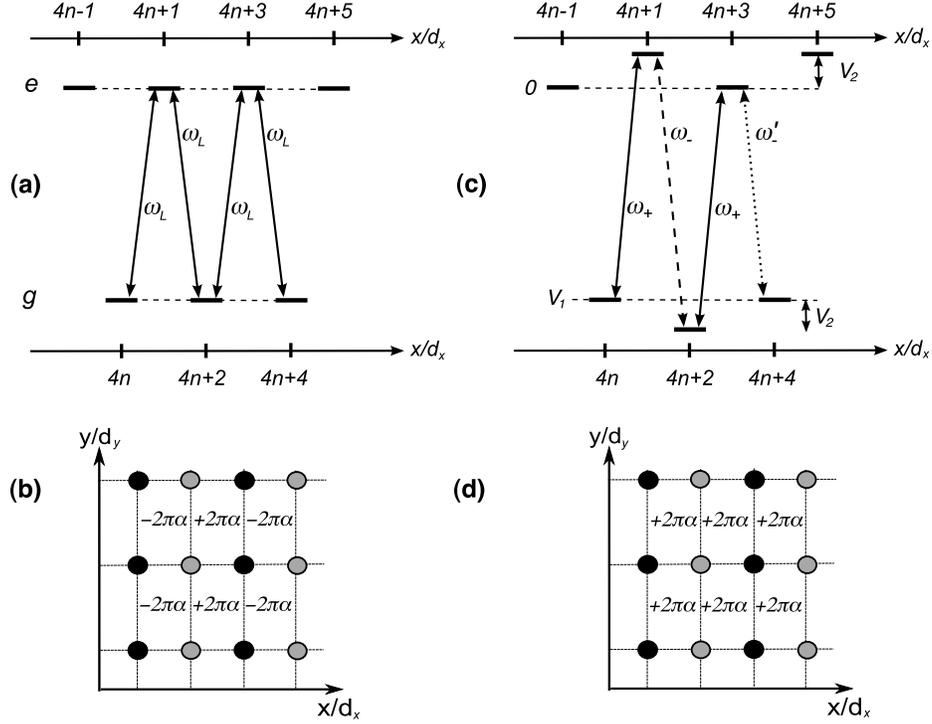}}
\caption{Laser schemes for realizing staggered ({\bf a,b}) and uniform  ({\bf c,d})  magnetic fields. {\bf (a)}: The figure shows the different on-site energies along $x$, in the case of a simple-period lattice. All transition frequencies are degenerate, so that a single laser beam at frequency $\omega_{L}$ is sufficient to couple all sites across the lattice. {\bf (b)}: Staggered effective magnetic field for a single coupling laser propagating along $y$ with degenerate transition frequencies as shown in {\bf (a)}. {\bf (c)}: In a configuration where an additional, doubly-periodic shift of the on-site energies is introduced, the degeneracy between the transition frequencies is lifted. {\bf (d)}: With several dedicated laser beams ($\omega_{+}$ propagating along $y$, $\omega_{-}$ and $\omega'_{-}$ propagating along $-y$), one obtains a ``rectified'' uniform magnetic field.}
\label{fig:YbSuperLattice}
\end{figure*}

We first consider the simplest experimental arrangement, with a spin-dependent lattice as described above and a single coupling laser (Fig.~\ref{fig:YbSuperLattice}a). The two-component $g-e$ quantum gas, which is assumed to populate the lowest Bloch band only, is described by the Hamiltonian 
\begin{eqnarray} \nonumber
H &=& -J \sum_{n,m, n\, {\rm even}}~\Bigl(e^{i 2\pi\alpha m} \hat{c}^\dagger_{2n+1,m} \hat{c}_{2n,m}+e^{i 2\pi\alpha m} \hat{c}^\dagger_{2n-1,m} \hat{c}_{2n,m}+ {\rm h.c.}\Bigr)\\
&  &- J \sum_{n,m,\pm} ~ \Bigl( \hat{c}^\dagger_{n,m\pm1} \hat{c}_{n,m} + {\rm h.c.} \Bigr),
\label{eq:2comp}
\end{eqnarray}
when the lattice amplitudes are adjusted so that $J^{(y)}=J_{eg}^{(x)}=J$. As in \cite{jaksch2003a}, the operator $\hat{c}_{n,m}^\dagger$ creates an atom at site $(n,m)$, the parity of $n$ identifying the internal state unambiguously. Alternatively, the internal degrees of freedom can be eliminated by going to a dressed state basis, and the spatial wavefunctions associated with each dressed state obey the {\it same} Harper equation \cite{harper1955a,hofstadter1976a} as the one derived from (\ref{eq:hofstadter}). Hence, at least in the absence of interactions (which we assume through this paper), the internal state is merely a label that can be ignored in the analysis.

In the original equation (\ref{eq:hofstadter}), the phase picked up across each link has the same sign for a given link direction in {\it real} space$: e^{+i 2\pi\alpha m}$ in the $+x$ direction and $e^{-i 2\pi\alpha m}$ in the $-x$ direction. Here, we find in contrast that the phase is tied to the direction in {\it internal} space:  $e^{+i 2\pi\alpha m}$ for $g\rightarrow e$ transitions and $e^{-i 2\pi\alpha m}$ for $e\rightarrow g$ transitions. In real space, given the sublattices geometry, this corresponds to a phase that alternates in sign from an arbitrary unit cell (or ``plaquette'')  to the neighboring one along the $x$ axis. More precisely, consider a particle going clockwise around an elementary cell (figure \ref{fig:YbLevels}c). The phase picked up along the path $(2n,m-1)_{g}\rightarrow (2n,m)_{g} \rightarrow (2n+1,m)_{e} \rightarrow (2n+1,m-1)_{e} \rightarrow (2n,m-1)_{ g}$ is the product of the two factors $e^{i2\pi\alpha m}e^{-i2\pi\alpha (m-1)}=e^{i2\pi\alpha}$. Thanks to the phase factor $\propto e^{i {\bf q}\cdot {\bf r}_{g}}$, a particle making a loop around a plaquette acquires a non-zero global phase factor, reflecting the non-zero ``magnetic flux'' through the cell. However, for the adjacent plaquette, the path (also with clockwise orientation)  $(2n+1,m-1)_{e}\rightarrow (2n+1,m)_{e} \rightarrow (2n+2,m)_{g} \rightarrow (2n+2,m-1)_{g} \rightarrow (2n+1,m-1)_{e} $ leads to a phase factor $e^{-i2\pi\alpha}$ opposite to the previous one. One then achieves in this way a {\it staggered} magnetic field, with a phase per plaquette $\pm2\pi\alpha$ constant on a given column and changing sign from one column to the next (Fig. \ref{fig:YbSuperLattice}b). 

The case $\alpha=\frac{1}{2}$ is peculiar, since in this case a staggered or a uniform flux both correspond to a tunneling phase $e^{\pm i 2\pi\alpha m}=(-1)^m$ alternating between adjacent links depending on the parity of $m$. The Hamiltonian remains periodic, with a unit cell $d_{x} \times 2 d_{y}$ and two sites per unit cell. The energy spectrum is given by
\begin{equation}
E(k_x,k_y)=\pm 2 J \sqrt{\cos^2(k_x d_{x})+\cos^2(k_y d_{y}) },
\end{equation} 
where the Bloch vector $(k_x,k_y)$ sits in the first Brillouin zone $]-\pi/d_{x},\pi/d_{x}]\times ]-\pi/2d_{y},\pi/2d_{y}]$. We note that this dispersion relation gives rise to two `Dirac points" for $(k_x,k_y)=(\pm\pi/2d_x, \pi/2d_y)$ around which the dispersion relation is linear \cite{lim2008a,Hou2009a} (see also \cite{zhao2006a,zhu2007a,wu2008a,lee2009a,goldman2009b} for discussions of the Dirac points occurring in different settings, such as hexagonal lattices or non-Abelian gauge fields). The physical consequences of the existence of these Dirac points are similar to those found in the case of Graphene \cite{neto2009a}. However, their topological robustness \cite{montambaux2009a} might be affected by small deviations from $\alpha=1/2$ that break the two-site periodicity.

For $\alpha\neq \frac{1}{2}$, the staggered value of the phase per plaquette leads to a Hamiltonian that is different from (\ref{eq:hofstadter}), and which, in fact, does not exhibit any structure reminiscent of Hofstader's butterfly \cite{wang2006a}. This can be linked to a general argument due to Haldane \cite{haldane1988a}, emphasizing that breaking time-reversal invariance symmetry is the key factor to generate systems with properties similar to (integer) quantum Hall phases. The situation considered up to this point, where the same laser beam drives all $g \rightarrow e$ transitions, is invariant with respect to time-reversal. We describe in the next section a mean to break this invariance and achieve a flux with a constant sign on all plaquettes.

\section{State-dependent superlattice: Harper Hamiltonian for arbitrary $\alpha$}

\subsection{Flux rectification}\label{sec:rect}

To ensure an identical phase $2\pi \alpha$ per plaquette, instead of the staggered values $\pm 2\pi \alpha$, we consider a situation where the on-site energies are modulated spatially along $x$ with a period equal to twice the lattice spacing, as depicted in Fig.~\ref{fig:YbSuperLattice}c (the $y$ lattice remains identical for both states and is ignored in the following). Assume that the on-site energies are modulated according to 
\begin{eqnarray}
E_{g}-V_{1},~\mbox{for} ~x/d_{x}= 4n,\nonumber\\
E_{e}+V_{2},~\mbox{for} ~x/d_{x}= 4n+1,\nonumber\\
E_{g}-V_{1}-V_{2},~\mbox{for}~ x/d_{x}= 4n+2,\nonumber\\
E_{e},~\mbox{for} ~x/d_{x}= 4n+3.\nonumber
\end{eqnarray} 
Here the $E_{\alpha}$ ($\alpha=e,g$) denote the internal energies in free space, where $V_{1}$ is the amplitude of the fundamental lattice with period $d_{x}$ and where $V_{2} \ll V_{1}$ is the amplitude of the modulation with period $2 d_{x}$. In this potential landscape, the resonance frequencies for transitions linking neighboring sites become non-degenerate  (Fig.~\ref{fig:YbSuperLattice}c),
\begin{eqnarray}
\omega_{+}=\omega_{0}+(V_{1}+V_{2})/\hbar &~\mbox{for} ~~\vert g; 4n,m\rangle \rightarrow \vert e; 4n+1,m\rangle,\nonumber\\
\omega_{-}=\omega_{0}+(V_{1}+2V_{2})/\hbar\quad &~\mbox{for}~~ \vert e; 4n+1,m\rangle \rightarrow \vert g; 4n+2,m\rangle,\nonumber\\
\omega_{+}=\omega_{0}+(V_{1}+V_{2})/\hbar &~\mbox{for}~~ \vert g; 4n+2,m\rangle \rightarrow \vert e; 4n+3,m\rangle,\nonumber\\
\omega_{-}'=\omega_{0}+V_{1}/\hbar &~\mbox{for} ~~\vert e; 4n+3,m\rangle \rightarrow \vert g; 4n+4,m\rangle,\nonumber
\end{eqnarray} 
with $\hbar\omega_{0}=E_{e}-E_{g}$.

To the state-dependent lattice, one thus applies three coupling lasers propagating along $y$ with frequencies $\omega_{+}$, $\omega_{-}$ and $\omega_{-}'$. The laser at frequency  $\omega_{+}$ is chosen with a wavevector ${\bf k}_L$, and the lasers at frequencies $\omega_{-}$ and $\omega_{-}'$ with the opposite wavevector $-{\bf k}_L$. The alternance of the wavevectors compensates the alternance of the plaquette phases,  thereby ``rectifying'' the staggered flux obtained in a single-frequency configuration. If we neglect off-resonant transitions, an analysis similar to that made in the previous section confirms that each plaquette now acquires the same phase factor $e^{i2\pi\alpha}$, and that the system is indeed described by the Harper Hamiltonian (\ref{eq:hofstadter}).

\subsection{Optical superlattice}

To realize such a modulation in practice, we propose to superimpose along the $x$ direction an additional lattice potential with a double spatial period $2\lambda_{\rm am}$, which can be written as $V''(x)=V_{g/e}\cos(\pi x/4d_{x}+ \varphi)^2$, where $\varphi$ is the relative phase between the fundamental and double period lattices. Such superlattice potentials have been demonstrated experimentally for quantum gases \cite{sebbystrabley2006a,foelling2007a}. This potential can be realized in practice by adding a second laser with frequency $c/2 \lambda_{\rm am}$ phase locked to the fundamental lattice to control the relative phase $\varphi$ \cite{foelling2007a}. The amplitudes $V_{g/e}$ depend on the internal state and are fixed by the initial choice of wavelength for the fundamental lattice. The analysis in Section~\ref{sec:rect} above (where $V_{g}=V_{e}$) can be generalized straightforwardly to this less symmetric potential, with four non-degenerate transition frequencies in the general case. Moreover, it is possible to reduce the number of required laser frequencies to three by using a particular value of the relative phase $\varphi$ that reproduces exactly the scheme described previously, with $V_{2}=V_{e}\cos \varphi -V_{g}\sin \varphi$\footnote{Practically, the fluctuations $\delta \varphi$ of $\varphi$ must be small enough to ensure that fluctuations of $V_{2}$ remain much smaller than the width of the $g-e$ resonance, set by the Rabi frequency $\Omega$, or $\delta \varphi \ll \hbar \Omega/V_{2}\lesssim 0.1$. This is well within present technological capabilities.}.

\subsection{Conditions of validity}

We now discuss the domain of validity of our model. 
\begin{itemize}
\item First, as in the previous sections, we demand that laser-assisted tunneling dominates over regular tunneling without change in internal state. For a superlattice potential in the tight-binding regime, regular tunneling is suppressed when $V_{2} \geq E_{\rm R}$, since $J_{gg}^{(x)},J_{ee}^{(x)} \ll E_{\rm R}$. 

\item Second, the ratio $\hbar\Omega/V_{2}$, with $\Omega$ the Rabi frequency characterizing the atom-laser coupling, must remain small to neglect off-resonant transitions and describe the system by an effective Harper model \cite{jaksch2003a}. In practice, one can use a conservative value $\hbar\Omega=V_{2}/10$.

\item Finally, the modulation $V_{2}$ must be small enough to avoid excitations to higher-lying Bloch bands by one of the coupling lasers. A reasonable choice is $V_{2} = \Delta/3$, where $\Delta$ is the energy gap between the ground and first excited bands for $V_{2}=0$. This guarantees that transitions mixing different bands are detuned by at least $\Delta/3$. For example, if we focus on the $\vert e; 4n+3,m\rangle \rightarrow \vert g; 4n+4,m\rangle$ transition, transitions linking the ground and first excited bands occur at frequencies $\omega_{-}'\pm\Delta/\hbar$. One can check that the detunings for the different coupling lasers are $\vert V_{2}\pm \Delta\vert, \vert 2 V_{2}\pm \Delta\vert, \Delta$, which are indeed $\geq \Delta/3$ for the choice $V_{2}=\Delta/3$. 
\end{itemize}

We note that in the tight-binding regime,  $\Delta$ is notably larger than $ E_{\rm R}$, which allows one to fulfill simultaneously all these conditions while maintaining a substantial value for $J_{eg}^{(x)}$. We summarize our considerations in Table~\ref{Table:SL}, where we give typical values for lattice parameters and laser-assisted tunneling matrix elements in the superlattice configuration. We note that  the values for $J_{eg}^{(x)}$ remain sizeable, being on the same order as the tunnel energies $J_{gg}^{(x)}$ in the $g$ sublattice ($\hbar\Omega=V_{\rm 2}=0$) at moderate depths ($\sim10~E_{\rm R}$). 
\begin{table}[ht!]
\centering
\begin{tabular}{ |c|c|c|c|}
\cline{1-3}
Depth  & Energy gap &  Laser-assisted  \\
$V_{1}/E_{\rm R}$ &  $\Delta/E_{\rm R}$  & tunneling $J_{eg}^{(x)}/E_{\rm R}$  \\
\hline \hline
10 & 4.7 &  $16\times10^{-3}$ \\
20 & 7.8 & $13 \times10^{-3}$\\
30 & 9.8 & $9\times10^{-3}$ \\
\hline \cline{1-3}
\end{tabular}
\caption{Characteristic tunnel energies (along the $x$ axis) achievable for realistic experimental parameters in the superlattice configuration. Here $J_{eg}$ denotes the laser-assisted coupling from one sublattice to the other. We used a Rabi frequency $\hbar\Omega=V_{2}/10=\Delta/30~$ (see text).} \label{Table:SL}
\end{table}

\subsection{Comparison with the Jaksch-Zoller proposal}

The scheme as described in this paper is inspired by the earlier proposal by Jaksch and Zoller \cite{jaksch2003a}. That proposal was designed for alkali atoms in spin-dependent optical lattices. In this context, $g$ and $e$ denoted two different hyperfine states in the ground state manifold, and the spin-dependent optical potential was obtained by exploiting the vector light-shift arising in a laser field with suitable polarization (see \cite{dupontroc1973a,mandel2003a}, for example). 

The first difference between our method and the JZ proposal comes from the choice of two-electron atoms, versus ``one-electron'' alkalis. To obtain a useful spin-dependent lattice for akalis demands in practice to tune the lattice laser between the $D_{1}$ and $D_{2}$ lines, relatively close to resonance. For light alkalis, such as $^6$Li or $^{40}$K, the fine splitting between the $D_{1}$ and $D_{2}$ transitions is rather small (less than $1~$nm), and spontaneous emission rates are too large to maintain the cloud in the nK regime. The most practical candidate seems to be $^{87}$Rb, which has only stable bosonic isotopes and would probably still suffer from spurious spontaneous emission due to the coupling (Raman) lasers or from the lattice lasers. The choice of Yb removes all these drawbacks, by using only far-off resonant lasers and a dissipation-free coupling transition, and by allowing to study bosonic and fermionic systems in the same setup.

The use of a superlattice to ensure the same flux for all plaquettes constitutes the second main difference of our proposal with respect to JZ's proposal. There, a strong ``tilting'' potential independent of the internal state was introduced in order to distinguish $\vert g; 2n,m\rangle  \rightarrow  \vert e; 2n\pm1,m\rangle$. Although this could be done in principle using a large static electric field or the light shift from an intense far-off resonant laser beam, it turns out to be rather challenging from an experimental point of view\footnote{The Stark shift from the electric field is proportional to $ \alpha \vert E \vert^2$, with $\alpha$ the static polarizability. Using a configuration where the cloud is offset from the center of the electrodes creating the field, one can create a linear potential of the form $\eta x/d_{x}$, with $\eta\sim\alpha U^2 d_{x}/D$ with $D$ the distance between the electrodes and with $U$ the applied electrostatic potential difference. For $^{87}$Rb and $D\sim1~$cm, we find that a site-to-site energy offset on the order of the energy gap $\Delta$, $\eta\sim\Delta/3\sim10~$kHz requires $U\sim 50~$kV, which is rather challenging to produce.}.  In addition, this tilting potential has to be almost perfectly linear over the extent of the cloud. Otherwise, the various transition frequencies would depend on space, spoiling the flux uniformity and the scalability of the method. We believe that the use of a superlattice optical potential will be a more flexible and more reliable tool that can be readily implemented experimentally.

\section{Role of an external trap potential}

\begin{figure*}
\centerline{\includegraphics[scale=1]{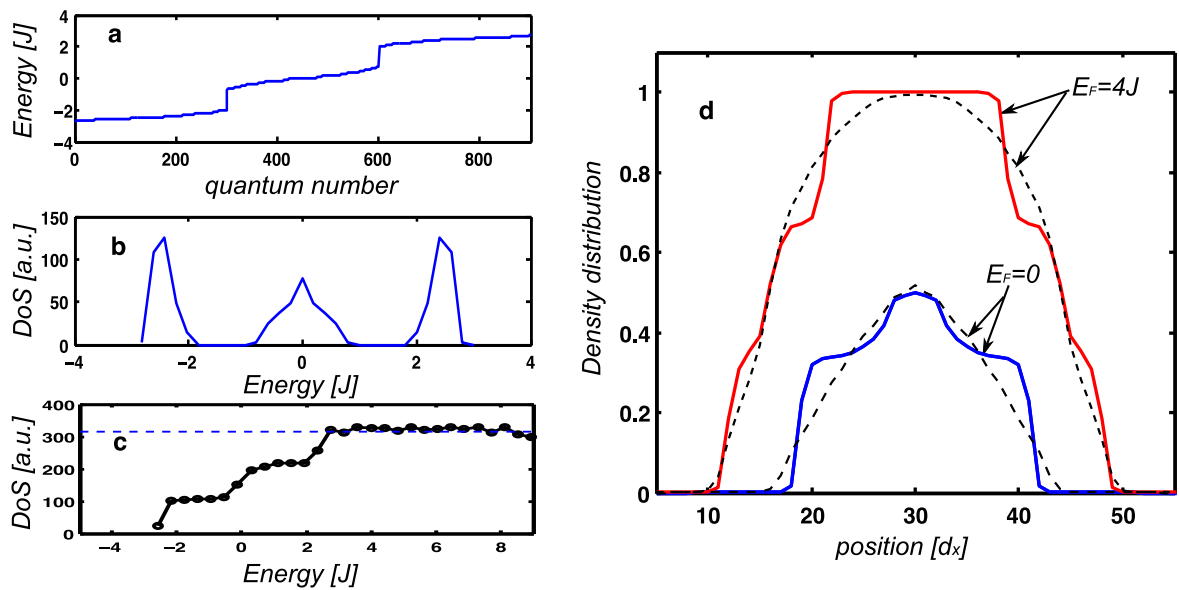}}
\caption{{\bf (a)}: Energy spectrum and {\bf (b)}: density of states (DoS) for the uniform model and $\alpha=1/3$. {\bf (c)}: density of states for the trapped model with $\alpha=1/3$ and confinement strength $\kappa/J=0.01$. The dashed line shows the expected density of states for $E\gg J$, irrespective of $\alpha$. {\bf (d)} Density profile for non-interacting fermions in a harmonic trap with $\kappa/J=0.01$, for a flux $\alpha=1/3$ (solid lines) and Fermi energies $E_{\rm F}=0$ (half-filled band) and $4J$ (filled band). The density profiles for $\alpha=0$ (dashed lines) are also shown for comparison. The calculation was performed on a $63\times63$ space grid.}
\label{fig:DoS}
\end{figure*}

We now discuss the properties of a system of non-interacting particles described by Eq.~(\ref{eq:hofstadter}) in order to identify observable consequences of the laser-induced gauge potential. For a uniform system without trapping potential, and for $\alpha=0$, the model reduces to a standard tight-binding Hamiltonian. The spectrum consists in an allowed energy band of width $8J$ centered at zero energy. For  the uniform Harper model, without trapping potential but with $\alpha=p/q\neq 0$ ($p$ and $q$ are integers), the initial Bloch band splits into $q$ magnetic sub-bands separated by well-defined energy gaps. This results in $q$ distinct peaks in the density of state (DoS-see Fig.~\ref{fig:DoS}a and b). In a practical experiment, an additional trapping potential is inevitably present in addition to the optical lattice. Hence the physical system is both finite and inhomogeneous. This can strongly affect the properties of the uniform Harper model given in Eq.~{\ref{eq:hofstadter}}, but also provide novel experimental signatures, as we discuss in this section. 

\subsection{Density of states in presence of an external trap}

We consider here the Harper model in Eq.~(\ref{eq:hofstadter}) with an additional isotropic trapping potential $V_{\rm trap}=\kappa (r/d)^2$. Let us recall that, with $\alpha=0$, one finds a quasi-continuum of allowed states in the combined lattice plus external trap, in striking contrast with the uniform case \cite{hooley2004a}. States with energy lower than $4J$ are qualitatively similar to the Bloch eigenstates of the uniform problem. These states are approximately contained in a disk of radius $R_{\rm eff}\sim\sqrt{J/\kappa}$. States with higher energy (which corresponds to the bandgap of the uniform system) are well-localized states offset from the cloud center \cite{ott2004a}, with an energy mostly determined by the local potential energy near the turning point. In two dimensions, the density of states saturates at an asymptotic value of $\pi/ \kappa$. 

We now address the case with $\alpha=p/q \neq 0$. The finiteness of the central region where tunneling is significant limits the relevant values of $q$ to $q<q_{\rm max}=R_{\rm eff}/d$. We consider here the case $\alpha=1/3$ for concreteness. Solving the full model numerically, we find that the trapping potential has a dominant effect on the DoS and little remains of the initial peak structure (compare Fig.~\ref{fig:DoS}b and Fig.~\ref{fig:DoS}c). This indicates that global observables (such as thermodynamic quantities) will not be very sensitive to the presence of the gauge potential. We can understand this behavior from a semiclassical argument. Indeed, in a local density approximation, the density of states in a generic power-law potential $\kappa (r/d)^\gamma$ can be approximated as
\begin{eqnarray}
\rho_{\rm trap}(E)=\sum_{n} \int \frac{d^{(2)} {\bf r} d^{(2)} {\bf k}}{(2\pi)^2} \delta\left( E -\epsilon_{n}({\bf k})-\kappa \left(\frac{r}{d} \right)^\gamma\right)
\end{eqnarray}
where $\epsilon_{n}({\bf k})=4 J \chi_{n}({\bf k})$ is the relation dispersion for the $n$th magnetic sub-band of the untrapped system. This can be simplified into
\begin{eqnarray}
\rho_{\rm trap}(E)=\frac{(4J)^{1-\frac{\gamma}{2}}}{\gamma \kappa}\sum_{n}\int_{\frac{E}{4J}>\chi_{n}({\bf k})} \frac{d^{(2) }{\bf k}}{S_{\rm BZ}} \left(\frac{E}{4J}-\chi_{n}({\bf k})\right)^{2-\gamma}.
\end{eqnarray}
Here $S_{\rm BZ}=(2\pi/d)^2$ is the surface of the Brillouin zone. For a harmonic potential ($\gamma=2$), the dependance on the reduced dispersion function $\chi$ is mild since it only arises from the upper bound of the integral: the peculiarities of the spectrum are essentially washed out by the spatial integration. This can be improved by using a steeper trap with $\gamma>2$, e.g. a quartic potential as demonstrated in \cite{bretin2004a}.

\subsection{Spatial distribution for non-interacting fermions}

In the case of non-interacting fermions in a gauge potential, a more sensitive observable is given by the particle density \cite{umucalilar2008a}, assuming it can be measured {\it in situ}. Indeed, still within a semiclassical approximation, the spatial density can be written as
\begin{eqnarray}\nonumber
n({\bf r})& = & \sum_{n}\int \frac{d^{(2)}{\bf k}}{(2\pi)^2}f_{\rm FD}\left[\epsilon_{n}({\bf k})+V_{\rm trap}(r)-E_{\rm F}\right],\end{eqnarray}
with $f_{\rm FD}$ the Fermi-Dirac distribution at $T=0$ and $E_{\rm F}$ the Fermi energy. We rewrite this as
\begin{eqnarray}\nonumber
n({\bf r})& = &  \sum_{n}\int \frac{d^{(2)}{\bf k}}{(2\pi)^2}\int_{-4J}^\infty dE~f_{\rm FD}\left[V_{\rm trap}+E-E_{\rm F}\right]~\delta\left[\epsilon_{n}({\bf k})-E\right]\\ &= &\int_{-4J}^\infty dE' f_{\rm FD}\left[V_{\rm trap}+E'-E_{\rm F}\right]~ \rho_{\rm hom}\left[E'\right],
\end{eqnarray}
where
\begin{eqnarray}
\rho_{\rm hom}\left[E\right]& = &  \sum_{n}\int \frac{d^{(2)}{\bf k}}{(2\pi)^2}~\delta\left[\epsilon_{n}({\bf k})-E\right]
\end{eqnarray}
is the density of state for the {\it homogeneous} system. For low enough temperatures, the clustering of energy levels in subbands in presence of a gauge potential will result in well-defined plateaux in the spatial density of the trapped cloud. These plateaux correspond to ``band-insulating'' regions, similar to that appearing for other incompressible systems in traps such as Mott insulators \cite{foelling2006a,campbell2006a}. We have verified this fact in our numerical calculations, as shown in Fig.~\ref{fig:DoS}d. We conclude that the in-trap density profile will provide a clear signature of the gauge potential in the case of fermionic atoms.

\subsection{Momentum distribution for non-interacting bosons}

\begin{figure*}[]
\centerline{\includegraphics[scale=1]{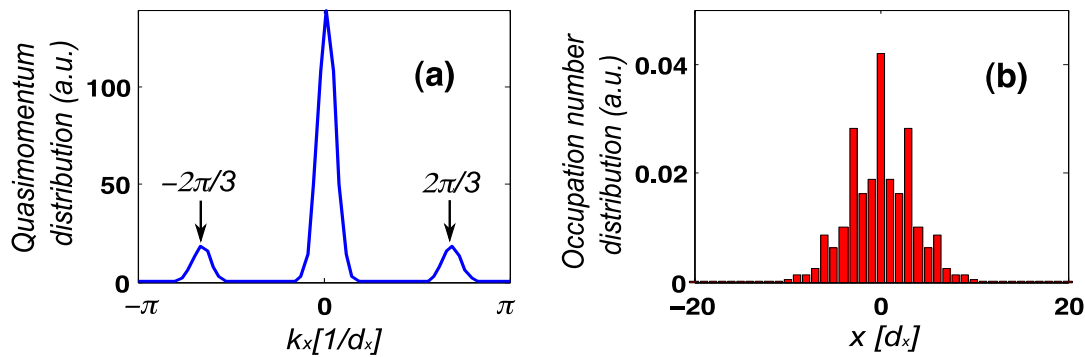}}
\caption{Typical momentum {\bf (a)} and spatial {\bf (b)} distributions in the ground state for non-interacting bosons and a flux $\alpha=1/3$. The presence of a gauge potential reveals itself in momentum components at $\pm 2 \pi/3 d$ in the former plot, and in the fine structures with period $3d$ in the latter. The confinement strength is $\kappa/J=0.005$.}
\label{fig:bosons}
\end{figure*}

With non-interacting bosons, such a signature will be absent, as the atoms will condense into the lowest single-particle states (for the purpose of this paper, we do not consider interactions). However, signatures of the gauge potential will appear in reciprocal momentum space, which can be probed using standard time-of-flight imaging. The dephasing term $e^{i2\pi\alpha m}$ generates quasi-momentum components at the various harmonics of the fundamental period $2\pi\alpha$. As a result, clear satellite peaks can be observed in the momentum distribution (see Fig.~\ref{fig:bosons}a), which will provide a direct experimental signature for bosonic species. We note in addition that the density distributions corresponding to these eigenstates shows periodic structures with period $q$ (see Fig.~\ref{fig:bosons}b), although this signature appears on a much finer scale than in the fermionic case (thus being more difficult to observe).

\section{Conclusion and perspectives}

In this article, we have discussed how to implement Abelian artificial gauge potentials for Yb (or alkaline-earth) atoms, a scheme valid for bosons and fermions alike. In contrast to the bosonic isotopes with zero total spin, fermionic isotopes have nuclear spin. This potentially enables to simulate non-Abelian gauge fields, along the lines proposed by Osterloh {\it et al.} \cite{osterloh2005a}. Let us focus for concreteness on the $^{171}$Yb isotope, with nuclear spin $I=\frac{1}{2}$. In the presence of a moderate magnetic field of a few tens of Gauss, the degeneracy of the $m_{I}=\pm\frac{1}{2}$ states within the $e$ and $g$ manifolds are lifted, giving in general four different internal transitions, $\pm \frac{1}{2} \rightarrow \pm \frac{1}{2}$ and $\pm \frac{1}{2} \rightarrow \mp \frac{1}{2}$, with different frequencies. At the expense of using more laser frequencies and state-dependent lattices along both $x$ and $y$ directions, one can then arrange for the lasers to imprint different phases depending on the internal state or generate arbitrary rotations in internal space. As shown in \cite{osterloh2005a}, a straighforward generalization of the Abelian scheme then allows to engineer {\it non-Abelian} gauge potentials. Instead of a simple phase, hopping from sublattice $g$ to $e$ then corresponds to a rotation in internal space generated by non-commuting matrices. Combined with interatomic interactions, an entirely new class of superfluid or strongly correlated systems becomes accessible with ultracold atoms
\cite{osterloh2005a,goldman2009b,goldman2007b,satija2008a,goldman2009a}. In a wider context, this technique can be used to emulate spin-orbit coupling, similar to the Rashba interaction $\propto \hat{\sigma}_{x}\cdot k_{y}-\hat{\sigma}_{y}\cdot k_{x}$ in planar semiconductors, or for the study of strongly interacting, ultracold fermionic matter in a non-Abelian background field, possibly relevant to high-energy physics.

We wish to thank the members of the ``Bose-Einstein condensates'' and ``Fermi gases'' group at LKB, G. Montambaux and G. Juzeliunas for helpful discussions, and A. Derevianko for a useful correspondence. This work was supported by ANR (Grant ANR-08-BLAN-65BOFL), IFRAF, the European Union (MIDAS project), and DARPA (OLE project). Laboratoire Kastler Brossel is unité mixte de recherche (UMR) n° 8552 of the CNRS.


\section*{References}

\end{document}